\documentclass[12pt,notoc]{JHEP3}

\usepackage{bm}
\usepackage{epsfig}
\usepackage{citesort}
\usepackage{graphicx}
\def\be{\begin{equation}}
  \def\ee{\end{equation}}
\def\bea{\begin{eqnarray}}
  \def\eea{\end{eqnarray}}

\preprint{CERN-PH-TH/2010-064, LAPTH-016/10, TTK-10-27}
\title{Cosmological parameters from large scale structure - geometric versus shape information}

\author{Jan Hamann, Steen Hannestad\\
  Department of Physics and Astronomy, University of Aarhus\\
  DK-8000 \AA rhus C, Denmark\\
  E-mail: \email{hamann@phys.au.dk}, \email{sth@phys.au.dk}}

\author{Julien Lesgourgues\\
  CERN, Theory Division, CH-1211 Geneva 23, Switzerland\\
  Institut de Th\'eorie des Ph\'enom\`enes Physiques, EPFL, CH-1015 Lausanne,
  Switzerland\\
  LAPTh (CNRS - Universit\'e de Savoie), BP 110, F-74941 Annecy-le-Vieux
  Cedex, France\\
  Email:\email{julien.lesgourgues@cern.ch}}

\author{Cornelius Rampf, Yvonne Y.~Y.~Wong\\
  Institut f\"{u}r Theoretische Teilchenphysik und Kosmologie, RWTH Aachen\\
  D-52056 Aachen, Germany\\
  E-mail: \email{rampf@physik.rwth-aachen.de}, \email{ywong@physik.rwth-aachen.de}}

\abstract{The matter power spectrum as derived from large scale
  structure (LSS) surveys contains two important and distinct pieces
  of information: an overall smooth shape and the imprint of baryon
  acoustic oscillations (BAO). We investigate the separate impact of
  these two types of information on cosmological parameter estimation
  for current data, and show that for the simplest cosmological
  models, the broad-band shape information currently contained in the
  SDSS DR7 halo power spectrum (HPS) is by far superseded by geometric
  information derived from the baryonic features.  An immediate
  corollary is that contrary to popular beliefs, the upper limit on
  the neutrino mass $m_\nu$ presently derived from LSS combined with
  cosmic microwave background (CMB) data does not in fact arise from
  the possible small-scale power suppression due to neutrino
  free-streaming, if we limit the model framework to minimal
  $\Lambda$CDM+$m_\nu$.  However, in more complicated models, such as
  those extended with extra light degrees of freedom and a dark energy
  equation of state parameter $w$ differing from $-1$, shape
  information becomes crucial for the resolution of parameter
  degeneracies.  This conclusion will remain true even when data from
  the Planck spacecraft are combined with SDSS DR7 data.  In the
  course of our analysis, we update both the BAO likelihood function
  by including an exact numerical calculation of the time of
  decoupling, as well as the HPS likelihood, by introducing a new
  dewiggling procedure that generalises the previous approach to
  models with an arbitrary sound horizon at decoupling.  These changes
  allow a consistent application of the BAO and HPS data sets to a
  much wider class of models, including the ones considered in this
  work.
  All the cases considered here are compatible with the conservative
  95\%-bounds $\sum m_{\nu} < 1.16\ {\rm eV}$, $N_{\rm eff}= 4.8 \pm
  2.0$.}

\begin{document}

\section{Introduction}
\label{sec:introduction}

Our best source of information about cosmological parameters at
present is the precision measurement of the cosmic microwave
background (CMB) anisotropies by the Wilkinson Microwave Anisotropy
Probe (WMAP)~\cite{Komatsu:2010fb}.  However, except in the simplest
models, the CMB on its own does not provide very tight constraints on
specific model parameters because of parameter degeneracies.  One very
well known example is the bound on neutrino masses, which in the
simplest vanilla+$m_\nu$ model can be {\it significantly} improved by
adding information extracted from surveys of the large scale structure
(LSS) distribution.  Moreover, if the model space is extended,
degeneracies with other parameters such as the dark energy equation of
state parameter quickly deteriorate the neutrino mass bound from CMB
data alone.  In such cases it is necessary to appeal to other
cosmological probes (e.g., LSS, Type Ia supernov\ae) in order to
alleviate the degeneracies.

In many recent analyses (e.g., \cite{Komatsu:2010fb}), the only
information from LSS surveys employed in the parameter estimation
pipeline is the length scale associated with the baryon acoustic
oscillation (BAO) peak in the two-point correlation function.  We call
this geometric information, since a known and measured length scale (a
``standard ruler'') allows for the determination of the angular
diameter distance to the object of interest simply via geometric
effects.  The common view is that the BAO length scale is a more
robust observable than the broad-band shape of the power spectrum
which may suffer from ill-understood nonlinear effects, such as
nonlinear clustering or redshift- and scale-dependent galaxy/halo
bias.%
\footnote{The turning point of the matter power spectrum in $k$ space
  corresponds to the comoving Hubble radius at the time of
  matter-radiation equality and in principle also constitutes a
  geometric measure.  However, since this length scale has yet to be
  measured, we prefer to consider the turning point as part of the
  broad-band shape of the power spectrum.}
Indeed, two recent studies of cosmological parameter constraints from
combining CMB information with either the BAO scale from SDSS
alone~\cite{Percival:2009xn}, or including {\it both} the broad-band
shape of the SDSS DR7 halo power spectrum and BAO
information~\cite{Reid:2009xm} find very similar parameter estimates
and uncertainties for the simplest vanilla model.  Somewhat
surprisingly, the same conclusion holds also when the vanilla model is
extended with a finite neutrino mass which should in principle be very
sensitive to the broad-band shape.  Thus circumstantial evidence seems
to suggest that in the simplest cosmological models, geometric
information from the BAO wiggles supersedes the information contained
in the broad-band shape of the matter power spectrum.

However, as we shall demonstrate, for more complicated models the
additional information contained in the broad-band shape of the matter
power spectrum can make a very substantial difference to parameter
inference .  For example, in models where the number of relativistic
degrees of freedom $N_{\rm eff}$ and the dark energy equation of state
parameter $w$ are added as free parameters, the difference in the
parameter uncertainties between including and excluding the power
spectrum shape information can be a factor of two or more (see also
Ref.~\cite{Biswas:2009ej} for a related discussion in the context of
CMB data and dark energy models).

The purpose of the present work is to clarify the roles of
``geometric'' and ``shape'' information extractable from the current
generation of LSS surveys, and to stress that in extended models
geometric information alone does not optimally exploit the available
data.  This will remain true even when CMB data from the Planck
spacecraft become available.  The paper is organised as follows: In
section~\ref{sec:analysis} we describe the specific cosmological
parameter space used, the data sets, as well as the analysis method.
We present our results in section~\ref{sec:results}, including a
forecast for Planck.  Our conclusions can be found in
section~\ref{sec:conclusions}.  Appendices~\ref{sec:appendix_hps} and
\ref{sec:appendix_bao} contain details of our methodology.

\section{Analysis}
\label{sec:analysis}

\subsection{Models}

We consider extensions of the minimal cosmological ({\it vanilla})
model, characterised by the free parameters listed in
Table~\ref{table:params}.  Flat priors are used on all parameters.%
\footnote{ Note that here, $N_{\rm eff}$ represents the number of
  massive neutrinos sharing a common mass $m_\nu$, while the neutrino
  fraction is defined as $f_\nu = {\omega_\nu}/{\omega_{\rm dm}}$ with
  $\omega_\nu=\Omega_\nu h^2 = [\sum m_{\nu}]/[93.14~{\rm eV}]$ and
  $\sum m_{\nu}=N_{\rm eff} m_{\nu}$.  This parameterisation of the
  neutrino sector is not very realistic from the particle physics
  point of view.  A better parameterisation might assign a free
  parameter $N_0$ to denote exclusively massless particle species,
  another parameter $N_{\rm m}$ for massive species, and $N_0+N_{\rm m}=N_{\rm
    eff}$.  However, the difference between the cosmological
  signatures of these various models is small given the sensitivity of
  the data, and introducing $N_{\rm eff}$ massive neutrinos (i.e.,
  setting $N_0=0$ and $N_{\rm m}=N_{\rm eff}$) as we do here makes for a
  much simpler analysis.}

\TABLE{ \centering
\begin{tabular}{|ccc|}
\hline
Parameter & Symbol & Prior range  \\
\hline
Baryon density & $\omega_{\rm b}$ & $0.005 \to 0.1$ \\
Dark matter density & $\omega_{\rm dm}$ & $0.01 \to 0.99$  \\
Hubble parameter & $h$ &  $0.4 \to 1.0$   \\
Reionisation optical depth & $\tau$ & $0.01 \to 0.8$ \\
Amplitude of scalar spectrum @ $k = 0.05\ {\rm Mpc}^{-1}$ & $\log \left[ 10^{10} A_{\rm S}\right]$ & $2.7 \to 4$ \\
Scalar spectral index & $n_{\rm S}$ & $0.5 \to 1.5$ \\
\hline
Neutrino fraction & $f_\nu$ & $0 \to 0.2$ \\
Effective number of radiation degrees of freedom & $N_{\rm eff}$ & $0 \to 30$ \\
Dark energy equation of state parameter & $w$ & $ -3 \to 0$ \\
\hline
\end{tabular}
\caption{\label{table:params} Cosmological parameters and prior
  ranges. The first six parameters constitute the well-known vanilla
  model.}
}

\subsection{Data}

Our main focus in this work is the halo power spectrum (HPS)
constructed from the luminous red galaxy sample of the seventh data
release of the Sloan Digital Sky Survey (SDSS-DR7)~\cite{Reid:2009xm}.
The full HPS data set consists of 45 data points, covering wavenumbers
from $k_{\rm min} = 0.02\ h {\rm Mpc}^{-1}$ to $k_{\rm max} = 0.2\ h
{\rm Mpc}^{-1}$ (where $k_{\rm min}$ and $k_{\rm max}$ denote the
wavenumber at which the the window functions of the first and last
data point have their maximum).  We shall use several subsets of the
full HPS data, taking into account only the first $5 X$ data points
(where $X$ is a natural number ranging from 1 to 9) -- this
corresponds to a short-scale cutoff between $k_{\rm max} = 0.04\ h
{\rm Mpc}^{-1}$ and $k_{\rm max} = 0.2\ h {\rm Mpc}^{-1}$, in steps of
$\Delta k_{\rm max} = 0.02\ h {\rm Mpc}^{-1}$.

Our goal is to disentangle the effects of shape and geometrical
information on cosmological parameter constraints.  To this end we
perform the following parameter fits:
\begin{itemize}
\item[1.]{We fit the HPS of Ref.~\cite{Reid:2009xm} as per usual, using a
    properly smeared power spectrum to model nonlinear mode-coupling.
    The smearing procedure requires that we supply a smooth,
    featureless (no-wiggle) power spectrum, which we construct using a
    new discrete spectral analysis technique (in contrast to the
    interpolation method used in Refs.~\cite{Reid:2009xm,Reid:2009nq}, which is, 
 strictly speaking, not applicable in extended cosmological models).  See
    Appendix~\ref{sec:appendix_hps} for details.  The result of this
    fit will contain {\it both} shape and geometric information.}
\item[2.]{We fit the no-wiggle spectrum alone to Reid {\it et al}.'s HPS data.  This
    fit singles out the broad-band shape information.}
\item[3.]{We use the the measurement of the baryon acoustic oscillation
    (BAO) scale obtained from SDSS-DR7~\cite{Percival:2009xn}, which
    represents solely the geometric information contained in the
    galaxy survey data.  Since the acoustic scale depends in principle
    on such parameters as $N_{\rm eff}$, care needs to be taken when
    evaluating the BAO likelihood.  We refer the reader to
    Appendix~\ref{sec:appendix_bao} for details.}
\end{itemize}
Since large scale structure data by themselves are not able to break
all the parameter degeneracies of our cosmological models, we complement
the power spectrum/BAO data with a compilation of measurements of the
CMB temperature and polarisation anisotropies, consisting of WMAP
7-year~\cite{Larson:2010gs},
ACBAR~\cite{Reichardt:2008ay}, BICEP~\cite{Chiang:2009xsa} and
QuAD~\cite{Brown:2009uy} data, as well as the HST constraint on the Hubble
parameter~\cite{Riess:2009pu}.  We avoid redundancies between the CMB
data sets in the same way as in Ref.~\cite{Finelli:2009bs}.

We employ a modified version of the Markov-chain Monte Carlo code
\texttt{CosmoMC}~\cite{Lewis:2002ah} to construct the posterior
probability density of the free model parameters.  For each combination
of model and data, we generate eight chains in parallel and monitor
convergence with the Gelman-Rubin $R$-parameter \cite{GelRu}, imposing
a conservative convergence criterion of $R-1 < 0.03$.

\section{Results}
\label{sec:results}

\subsection{Vanilla+$f_\nu$+$N_{\rm eff}$, $w = -1$ \label{subsec:v+f+n}}

Let us assume the dark energy to be a cosmological constant for now.
In Figure~\ref{fig:omn} we plot the constraints on the cosmological
parameters that benefit the most from the addition of LSS data to our
basic CMB+HST data set ($\omega_{\rm dm}$, $N_{\rm eff}$ and $\sum
m_\nu$) as a function of $k_{\rm max}$.  These results give rise to
the following observations:

\begin{itemize}
\item[1.]{There is no significant trend in the parameter estimates as
    smaller scale data are added, which implies a reassuring absence
    of obvious inconsistencies between the cosmological model and the
    HPS likelihood.}
\item[2.]{The greatest improvement in parameter constraints is
    apparent around $k_{\rm max} \sim 0.08-0.1\ h{\rm Mpc}^{-1}$.
    This is consistent with the fact that the BAO bump in the
    correlation function is around $100\ h^{-1}{\rm
      Mpc}$~\cite{Eisenstein:2005su}.  Hence one requires $\Delta k >
    0.06\ h{\rm Mpc}^{-1}$ to observe a full oscillation in the power
    spectrum in order to be  sensitive to the geometric information
    contained in the BAO scale.}
\item[3.]{Adding data points beyond $k_{\rm max} \simeq 0.12\ h{\rm
      Mpc}^{-1}$ does not lead to any further improvement in the parameter
    constraints.  This effect can be attributed to the modelling of
    nonlinear effects on the spectrum.  First, the onset of smoothing
    of the BAO due to mode coupling at $k_{\rm max} \sim 0.1\ h{\rm
      Mpc}^{-1}$, which limits the amount of information that can be
    gained about the BAO scale at large $k$, and second, the
    marginalisation procedure that is meant to model residual
    uncertainties in the nonlinear distortions of the spectrum, which
    obscures any shape information contained in the large-$k$ power
    spectrum.}
\item[4.]{Parameter constraints from a fit of a featureless no-wiggle
    spectrum to the full HPS data which ignores any geometrical
    information do not show any improvement over those derived from an
    analysis without large scale structure data.}
\item[5.]{The BAO data lead to slightly tighter bounds on the
    parameters than the HPS data (see Table~\ref{table:numbers}).
    Along with point 4., this implies that the shape information does
    not contribute any relevant additional information in this model.}
\end{itemize}
\FIGURE[H!]{
\centering
\includegraphics[width=.8\textwidth]{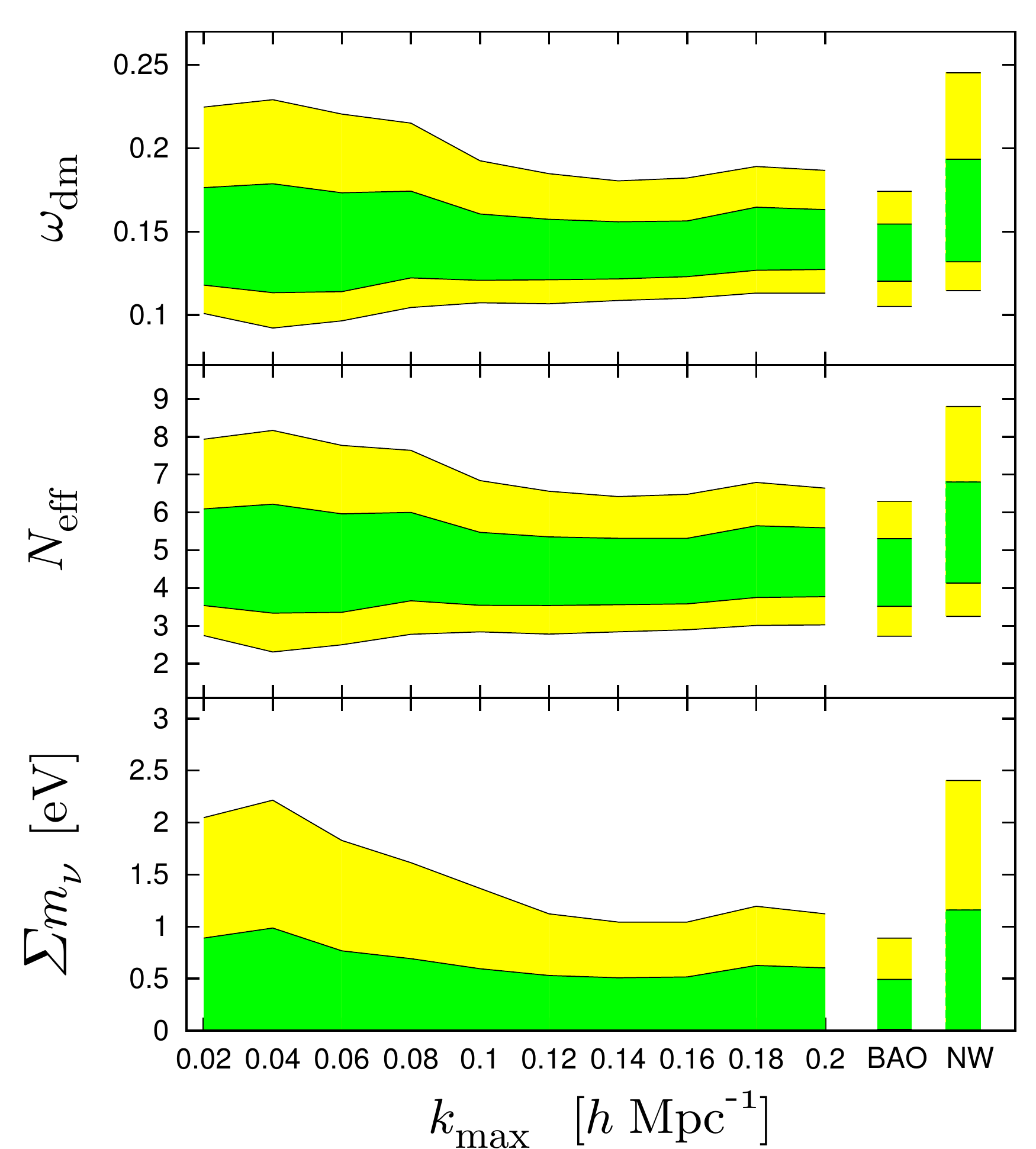}
\caption{Parameter constraints for the vanilla+$f_\nu$+$N_{\rm eff}$
  model as function of the maximum wavenumber $k_{\rm max}$,
  obtained for a combination of CMB+HST data with the halo power
  spectrum between wavenumbers of $k_{\rm min} = 0.02\ h{\rm
    Mpc}^{-1}$ and $k_{\rm max}$.  Bounds for $k_{\rm max} = 0.02\
  h{\rm Mpc}^{-1}$ correspond to results for CMB+HST only.  The
  marginalised minimal 68\%-credible intervals \cite{Hamann:2007pi}
  are marked in green, the 95\%-credible intervals in yellow.  For
  comparison we also show the constraints from CMB+HST+BAO data (denoted ``BAO'') and
  from a fit of a featureless no-wiggle spectrum to CMB+HST+HPS data (``NW'').}
\label{fig:omn}
}
The fact that BAO slightly outperforms HPS may seem somewhat
surprising at first glance, but there are several reasons that could
account for this phenomenon.  First, the BAO data make use of the full
SDSS galaxy sample instead of just the LRGs.  Second, the BAO data
constrains the BAO scale at two distinct redshifts, whereas the HPS
data as implemented at the moment only constrain the BAO scale at one
effective distance obtained from averaging over the mean redshifts of
the FAR, MID and NEAR LRG-subsamples.  Additionally, since the
geometric information is obtained through different analysis
pipelines, we cannot rule out the possibility that one of them may be
slightly more aggressive and thus yield tighter constraints.

On a side note, it is interesting to point out that due to a parameter
degeneracy between $N_{\rm eff}$ and $n_{\rm S}$ in this model the
favoured value for the spectral index is shifted to bluer tilts than
in the basic six-parameter vanilla model, with $n_{\rm S} = 0.981 \pm
0.014$ (@ 68\% c.l.) for CMB+HST+HPS. In other words, if $N_{\rm
  eff}>3$, the scale-invariant Harrison-Zel'dovich spectrum becomes
viable again, with $n_{\rm S} = 1$ corresponding to $N_{\rm eff}
\simeq 6$ (see the thin dotted black contour in the left panel of
Figure~\ref{fig:nswneff}).  Even though this would indicate a
non-standard cosmology, perhaps with a large amount of late-time
entropy production, it does show that with current data the inference
that $n_{\rm S} < 1$ is not completely robust.

\subsection{Vanilla+$f_\nu$+$N_{\rm eff}$+$w$}

Allowing the dark energy equation of state parameter $w$ to vary
introduces another direction which contributes to the geometric
degeneracy.  Consequently, one can expect a degradation in parameter
constraints compared to the $w = -1$ case considered in the previous
subsection.  The parameter most affected by this turns out to be the
spectral index, which is illustrated by the difference between the
thick black and thin dotted black contours in
Figure~\ref{fig:nswneff}.  The closer $w$ is to zero, the more power
at large angular scales in the CMB temperature spectrum will be
generated due to the late integrated Sachs-Wolfe effect.  This,
in turn, can be offset to fit present CMB data by removing large-scale
primordial power, i.e., going to a bluer spectral index.

\FIGURE[H!]{
\centering
\includegraphics[angle=270,width=.75\textwidth]{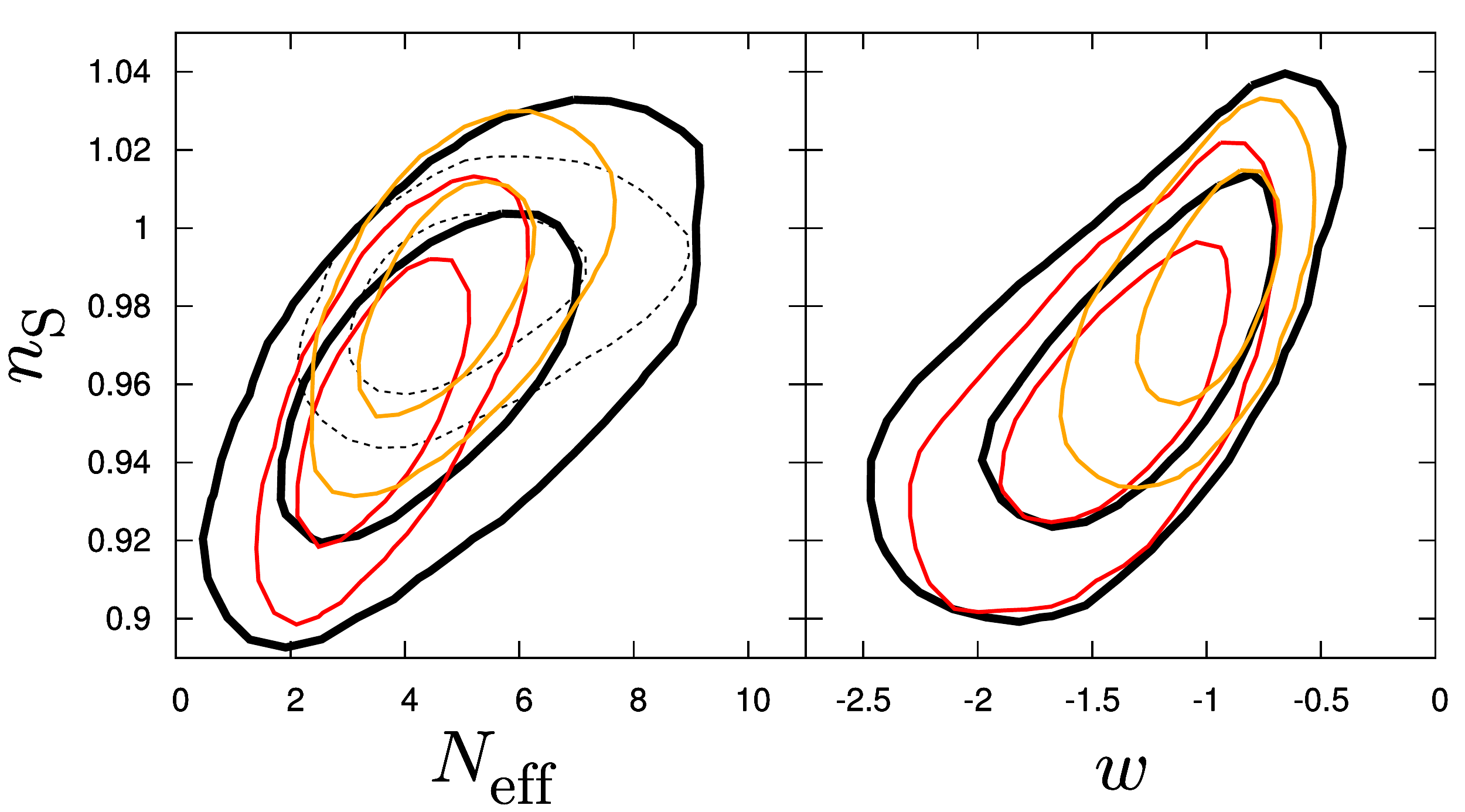}
\caption{This figure shows the 68\%- and 95\%-credible regions of the
  joint marginalised posterior in the ($n_{\rm S}$,$N_{\rm eff}$)- and
  ($n_{\rm S}$,$w$)-plane in the vanilla+$f_\nu$+$N_{\rm eff}$+$w$
  model for CMB+HST data (thick black lines), CMB+HST+BAO data (red
  lines) and CMB+HST+HPS data (orange lines).  The thin dotted black
  line in the left panel corresponds to the $w=-1$ case for CMB+HST
  and demonstrates how varying $w$ introduces a new degeneracy
  direction.
\label{fig:nswneff}}
}

\FIGURE[h!]{
\centering
\includegraphics[width=.8\textwidth]{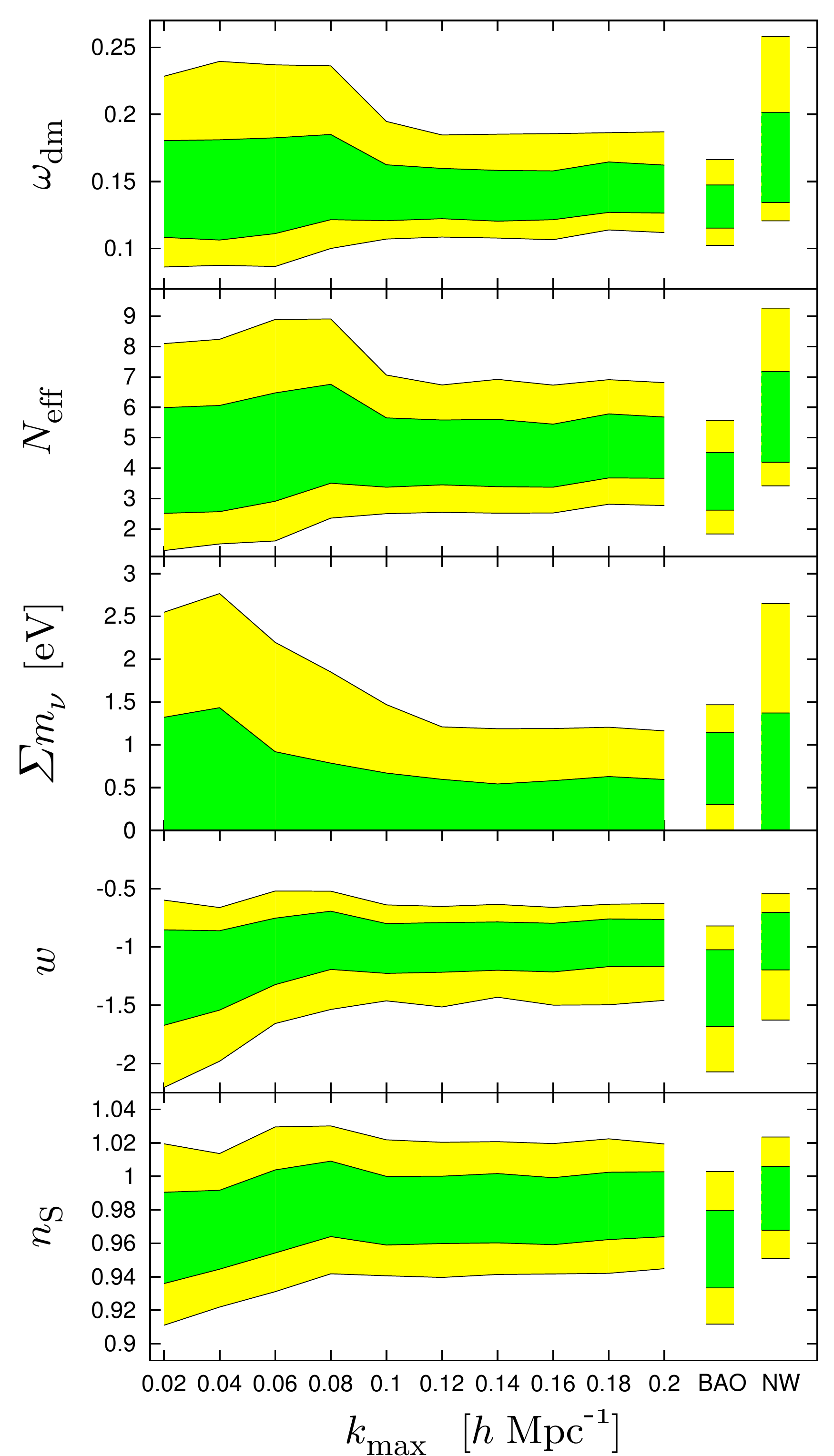}
\caption{Same as Figure~\ref{fig:omn} for the vanilla+$f_\nu$+$N_{\rm
    eff}$+$w$ model. \bigskip}
\label{fig:omnw}
}

From Figure~\ref{fig:omnw} it can be seen that the internal
consistency of the HPS and the sharp improvement in errors around
$k_{\rm max} \sim 0.1\ h{\rm Mpc}^{-1}$ are still present.  Notably
though, the BAO data alone are not able to break the ($n_{\rm
  S}$,$w$,$N_{\rm eff}$)-degeneracy present in the basic CMB+HST data.
Indeed, BAO are not sensitive to the spectral index.  In the previous
model with $w=-1$, $n_{\rm S}$ was reasonably well constrained by the
CMB data alone.

In the model at hand though, the CMB cannot resolve the ($n_{\rm
  S}$,$w$)-degeneracy. In order to break it, it becomes crucial to
include the shape information of the HPS spectrum.  This conclusion is
supported by the observation that a fit of the no-wiggle spectrum to
HPS data yields essentially the same bounds on $n_{\rm S}$ and $w$ as
the usual fit with full HPS data.  On the other hand, the best
constraints on parameters like $\omega_{\rm dm}$ still come from the
geometric rather than shape information.  So, a fit of the full HPS
data is necessary in order to get efficient constraints on this model.

\TABLE[h]{\centering
\begin{tabular}{|l|l|c|c|c|}
\hline
Model & Data set & $\sum m_\nu/{\rm eV}$ & $N_{\rm eff}$ & $w$\\
\hline
& CMB+HST & $< 2.05$ & $5.16^{+2.77}_{-2.42}$ & --\\
vanilla+$f_\nu$+$N_{\rm eff}$ & CMB+HST+BAO & $< 0.89$ & $4.47^{+1.82}_{-1.74}$ & --\\
& CMB+HST+HPS & $< 1.12$ & $4.78^{+1.86}_{-1.75}$ & --\\
\hline
& CMB+HST & $< 2.58$ & $4.68^{+3.72}_{-3.48}$ & $-1.33^{+0.77}_{-0.87}$ \\
vanilla+$f_\nu$+$N_{\rm eff}$+$w$ & CMB+HST+BAO & $< 1.47$ & $3.68^{+1.90}_{-1.84}$ & $-1.42^{+0.60}_{-0.65}$ \\
& CMB+HST+HPS & $< 1.16$ & $4.79^{+2.02}_{-2.02}$ & $-1.02^{+0.39}_{-0.44}$\\
\hline
\end{tabular}
\caption{\label{table:numbers}Means and limits of the 95\%-credible intervals
  for the non-vanilla parameters and various combinations of data
  sets.}}

\subsection{A forecast for Planck}
\TABLE[b!]{\centering
\begin{tabular}{|c|c|c|c|c|c|c|}
\hline
 & Planck & P+BAO & P+HPS & P+HST & P+HST+BAO & P+HST+HPS\\
\hline
$\omega_{\rm dm}$& 0.22 & 0.24 & 0.20 & 0.21  & 0.21  & 0.19 \\
$N_{\rm eff}$& 0.21 & 0.21 & 0.22 & 0.21 & 0.21 & 0.22 \\
$\sum m_\nu$& 0.68 & 0.81 & 0.44 & 0.67 & 0.73 & 0.44 \\
$w$& 2.14 & 1.16 & 0.72 & 0.74 & 0.76 & 0.55 \\
$n_{\rm S}$& 0.46 & 0.48 & 0.49 & 0.46 & 0.48 & 0.48 \\
\hline
\end{tabular}
\caption{\label{tab:plonk}Projected sensitivity of Planck data (P)
  combined with LSS data to selected parameters of the
  vanilla+$f_\nu$+$N_{\rm eff}$+$w$ model.  Given are the standard
  deviations of the marginalised posteriors, normalised to the values
  obtained with current CMB+HST+HPS data.  Note that just like for
  current CMB data, the addition of BAO data shifts the posterior
  towards larger neutrino masses, resulting in a two-tailed pdf with a
  correspondingly larger standard deviation -- this does not mean that
  the constraining power of Planck+BAO is worse than that of Planck
  alone.  The marginalised posteriors of all the other parameters are
  very close to two-tailed Gaussians, and do not suffer from this
  effect.}}
The conclusions one can draw about the usefulness of geometric and
shape information obviously depend not only on the large scale
structure data themselves, but also on the constraining power of the
``auxiliary'' data used in the analysis.  It is thus interesting to
ask whether one can expect any qualitative changes from improved
future cosmological measurements for the vanilla+$f_\nu$+$N_{\rm
  eff}$+$w$ model.  As an example we consider a forecast of simulated
fiducial temperature and polarisation data from the Planck
spacecraft~\cite{planck} using the method of
Ref.~\cite{Perotto:2006rj}.  Our results in Table~\ref{tab:plonk} show
that Planck data alone will suffice to constrain $\omega_{\rm dm}$,
$n_{\rm S}$ and $N_{\rm eff}$.  Nonetheless, the bounds on neutrino
mass and $w$ will still profit from the addition of large scale
structure data -- and significantly, the SDSS shape information will
remain important.

\section{Conclusions}
\label{sec:conclusions}

We have studied in detail how various subsets of the power spectrum
information from a large scale structure survey can be used to
constrain cosmological parameters.  At present an often used approach
is to restrict the LSS information to the geometric distance
information contained in the BAO peak.  For the minimal $\Lambda$CDM
model (with neutrino mass included) this does indeed provide exactly
the same constraint as the use of the full power spectrum data, and is
far superior to using a smoothed no-wiggle power spectrum which
contains only shape information. 
This indicates that the neutrino mass is currently more strongly
constrained by its effect on the background
evolution~\cite{Lesgourgues:2006nd}, and the contribution of present
LSS data consists mainly in alleviating the geometrical degeneracy
with $h$ and $\Omega_{\rm m}$ \cite{Komatsu:2008hk,Thomas:2009ae},
rather than constraining the possible small-scale power suppression in
the large scale matter power spectrum due to free-streaming.

However, this simple picture changes when more complex cosmological
models are studied.\footnote{We should point out a small caveat here:
  the usefulness of the LSS shape information depends not only on the
  cosmological model under consideration, but also on the combination
  of data sets used in the analysis.  For example, Reid et
  al.~\cite{Reid:2009xm} find that in a vanilla+$N_{\rm eff}$ model,
  WMAP5+HPS yields much better constraints on $N_{\rm eff}$ than
  WMAP5+BAO.  However, in this model the LSS shape information loses
  its usefulness as soon as one adds the HST constraint, which breaks
  the ($H_0$-$N_{\rm eff}$)-degeneracy more efficiently -- leading to
  conclusions similar to those found in our
  subsection~\ref{subsec:v+f+n}.}  As an example we have tested a
model with a variable number of neutrino species, and a dark energy
equation of state, $w$, different from -1.  In this model some
parameters are still as well constrained by BAO alone as by the full
power spectrum.  This is true for example for the number of neutrino
species, which is mainly probed by CMB data, and the dark matter
density which is highly sensitive to the position of the BAO peak.
However, for other parameters such as $\sum m_\nu$, $w$ and $n_{\rm
  S}$, there is additional information in the shape of the power
spectrum which is crucial for constraining these parameters.  For
example the upper 95\% bound on neutrino mass goes from $1.47\ {\rm
  eV}$ to $1.16\ {\rm eV}$ when BAO information is replaced with the
full halo power spectrum.

However, even in this model, the entire shape information of the HPS
is contained in the data points at wavenumbers smaller than $0.12\ h
{\rm Mpc}^{-1}$, the higher-$k$ information being diluted due to
uncertainties in nonlinear modelling.  In other words, due to our
ignorance of the processes governing the power spectrum at smaller
scales, we basically lose almost half of the available data points
(the half that is less subject to sample variance at that!).  Clearly,
a better understanding of the mildly nonlinear physics at these scales
would be highly desirable.

Needless to say, our conclusions do not alter the fact that in the
future, when better data from galaxy, cluster, weak lensing or 21cm
surveys will be available, the best way to probe the neutrino mass
will be through the information contained in the shape and the
scale-dependent growth factor of the large scale structure power
spectrum~\cite{Song:2004tg,Lesgourgues:2004ps,Wang:2005vr,Lesgourgues:2006nd,Hannestad:2007cp,Vikhlinin:2008ym,Pritchard:2009zz}.

In this paper we have also presented a new method for separating the
geometric BAO information from the shape information in the no-wiggle
spectrum for more complex models than previously studied. The method
is based on removing the oscillating part of the power spectrum by use
of a fast sine transform and then removing the BAO peak by smoothing
the resulting ``correlation function''.  Finally, the smoothed function
is transformed back to provide the no-wiggle power spectrum. The
method has been demonstrated to be extremely fast and robust to even
radical changes in the cosmological model, making it easy and safe to
implement in \texttt{CosmoMC}.  Along the same lines we have also
implemented a version of the SDSS BAO likelihood code which allows for
models in which the sound horizon is modified compared to the
$\Lambda$CDM model with $N_{\rm eff}=3$.

In addition to constraints using current data we have also performed
an estimate of how the SDSS measurements can be used to improve the
Planck constraints on some parameters in extended models.  Most
parameters will be so well determined by Planck that little can be gained
from adding the SDSS data in any form. However, with $\sum m_\nu$ and
$w$ the situation is different. With these parameters the SDSS data
can lead to very significant improvements in sensitivity. Furthermore,
we have also shown that even for Planck the SDSS halo power spectrum
contains important information beyond what is in the geometric BAO
data - for both $\sum m_\nu$ and $w$ the shape information can improve
the sensitivity by 30-40\%. 

As shown in Ref.~\cite{Rassat:2008ja}, in future large scale structure
surveys the relative impact of the shape information is expected to
increase, so extracting the full power spectrum information and at the
same time improving the theoretical modeling of small-scale
perturbations remains a crucial goal.


\acknowledgments{We thank Beth Reid and Licia Verde for interesting
  discussions.  JH acknowledges the support of a Feodor
  Lynen-fellowship of the Alexander von Humboldt foundation.  JH, JL
  and SH also acknowledge support from the EU 6th Framework Marie
  Curie Research and Training network `UniverseNet'
  (MRTN-CT-2006-035863). We used computing resources from the Danish
  Center for Scientific Computing (DCSC) and from the MUST cluster at
  LAPP, Annecy (CNRS \& Universit\'e de Savoie). WMAP7 data is made
  available through the Legacy Archive for Microwave Background Data
  Analysis (LAMBDA), supported by the NASA Office of Space Science.}

\appendix

\section{Halo power spectrum data in extended models}
\label{sec:appendix_hps}

Once perturbations pass into the nonlinear regime, mode coupling will
set in and fluctuations of different wavenumbers no longer evolve
independently.  As a consequence, any feature in the matter power
spectrum, most notably the baryon acoustic oscillations, will be
washed out beyond $k > 0.1\ h {\rm Mpc}^{-1}$.  This aspect of
nonlinear evolution can be modelled by considering the smeared power
spectrum~\cite{Eisenstein:2006nj}, a scale-weighted average of the
linear power spectrum $\mathcal{P}_{\rm lin}(k)$ and a featureless
{\em no-wiggle} spectrum $\mathcal{P}_{\rm nw}(k)$,
\begin{equation}
  \mathcal{P}_{\rm smear}(k) = \mathcal{P}_{\rm lin}(k) \, \exp \left(- \frac{ k^2
      \sigma^2}{2} \right) + \mathcal{P}_{\rm nw}(k) \, \left( 1 - \exp \left(-
      \frac{k^2 \sigma^2}{2} \right) \right),
\end{equation}
with a redshift-dependent smoothing scale $\sigma \sim 85\ h^{-1}{\rm
  Mpc}$ (at $z=0$) calibrated by simulations~\cite{Reid:2009xm}.  For
the no-wiggle portion of the spectrum, which taken by itself
encapsulates the shape information in the data, two approaches have
been used in the recent literature: a semi-analytic fitting formula,
originally introduced in Ref.~\cite{Eisenstein:1997ik}, and a
cubic-spline interpolation of the linear power-spectrum given some
fixed nodes.
 
While fine for standard $\Lambda$CDM cosmology, the semi-analytic
formula introduced in Ref.~\cite{Eisenstein:1997ik} {\it per se} does
not describe cosmological models extended with massive neutrinos, a
non-trivial equation of state for the dark energy, or non-standard
relativistic degrees of freedom, to name a few.  Extensions to the
fitting formula of Ref.~\cite{Eisenstein:1997ik} to include a non-zero
$f_\nu$ and $w \neq -1$ have been investigated in
Ref.~\cite{Eisenstein:1997jh} and \cite{Kiakotou:2007pz} respectively.
The second approach, the cubic-spline interpolation method of
Ref.~\cite{Reid:2009xm}, bypasses completely the use of fitting
formulae.  Instead, it removes the BAO by singling out a reference set
of oscillation nodes (corresponding to the WMAP5 best-fit vanilla
model), which are then interpolated using a cubic spline.  This method
is correct as long as the chosen interpolation points coincide with
the actual nodes of the BAO for a given cosmological model.  However,
in any model that tampers with the sound horizon relative to the
reference case, e.g., by allowing $N_{\rm eff}$ to vary, the
interpolation nodes and the actual nodes of the BAO can shift out of
phase, thereby resulting in an insufficient removal of the baryon
wiggles.  We demonstrate an extreme example in
Fig.~\ref{fig:badnowiggle}.  

\FIGURE[h!]{ \centering
\includegraphics[angle=270,width=.6\textwidth]{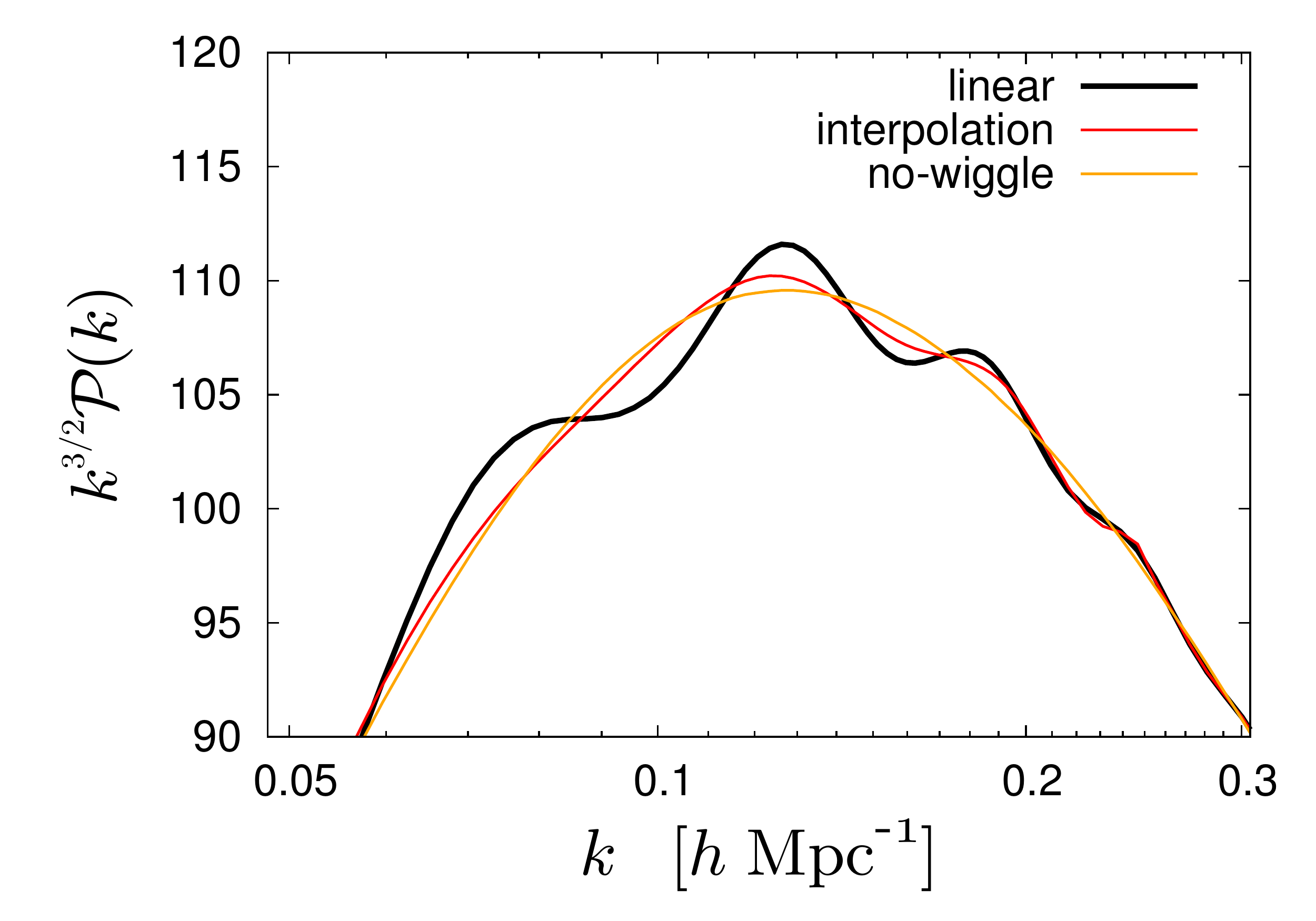}
\caption{Example of a model for which the interpolation method of Ref.~\cite{Reid:2009xm} fails
  to yield a properly de-wiggled spectrum.  Parameter values are
  $\omega_{\rm b} = 0.0223$, $\omega_{\rm dm} = 0.25$ and $N_{\rm eff}
  = 10$.  The orange line is the no-wiggle power spectrum constructed
  with the spectral analysis method described below.}
\label{fig:badnowiggle}
}
 
Clearly, to ensure the nonlinear mode coupling is properly modelled in
extended cosmological models, a more reliable way of constructing a
smooth no-wiggle spectrum is needed.  In this work, we have explored
three different alternative ways to produce a no-wiggle spectrum: (i)
a method based on a discrete spectral analysis of the power spectrum
in log-space, (ii) a fourth order polynomial fit in $\log(k)$, and
(iii) a semi-analytic fitting formula with $f_\nu$, $N_{\rm eff}$ and
$w$ as free parameters built on the works of
Refs.~\cite{Eisenstein:1997ik,Eisenstein:1997jh,Kiakotou:2007pz}.
These are described in detail in the following subsections.  The
spectral method (i) is arguably by far the most elegant and generally
applicable of the three alternative methods proposed here; all
parameter estimation results presented in this work have been obtained
using this approach.

\subsection{Discrete spectral analysis}

This approach is similar to considering the correlation function
\begin{equation}
  \label{eq:corrfcn}
  \xi(r) = \frac{1}{\sqrt{2 \pi}} \int {\rm d}k \; k \mathcal{P}(k) \,
  \frac{\sin (kr)}{kr},
\end{equation}
in which the oscillations in $\mathcal{P}(k)$ appear as a
bump~\cite{Eisenstein:2005su}.  Numerically, it is by far easier to
identify a bump in an otherwise smooth function than it is to, e.g.,
find the zeros of the BAO in the power spectrum.  Once the BAO bump
has been identified and removed, the ``de-bumped'' correlation
function can then be inverse Fourier-transformed to obtain a 
no-wiggle spectrum.

In practice, it turns out to be more efficient to use a discrete
Fourier transform (more precisely, a fast sine transform~\cite{NR})
instead of evaluating the integral~(\ref{eq:corrfcn}).  We have
implemented the following algorithm to construct the no-wiggle
spectrum:
\begin{itemize}
\item[1.]{Sample $\ln (k \mathcal{P}_{\rm lin}(k))$ in $2^n$ points,
    equidistant in $k$.}
\item[2.]{Do a fast sine transform of the array constructed in 1.}
\item[3.]{Interpolate the odd and even entries of the resulting array
    using a cubic spline~\cite{NR}.}
\item[4.]{Identify the baryonic bumps by determining where the second
    derivatives of the splines starts becoming large.}
\item[5.]{Cut out the points corresponding to the baryonic bumps and
    fill in the gap by cubic spline interpolation.}
\item[6.]{Do a reverse fast sine transform to recover $\ln (k
    \mathcal{P}_{\rm nw}(k))$.}
\end{itemize}
For the models considered in this paper the effect on parameter
estimates is relatively small, but nevertheless noticeable.  Compared
to results obtained with the spectral smoothing method, the fixed-node
interpolation smoothing of Ref.~\cite{Reid:2009xm} overestimates the
uncertainty on $N_{\rm eff}$ and $n_{\rm S}$ by roughly $10\%$, and
the error on $w$ by about $25\%$ in the vanilla+$f_\nu$+$N_{\rm
  eff}$+$w$ model.  There is, however, no significant bias on the
estimates of parameter means.

\subsection{Fourth order polynomial fit}

The polynomial fit method consists of replacing the logarithm of the
power spectrum $\ln \mathcal{P}_{\rm lin}$ by a fourth-order
polynomial in $(\ln k)$ within the range $[k_1, k_2]$, where $k_1$ is
the turn-over scale of the spectrum (found automatically for each
model), and $k_2$ is a scale fixed by the user (here, we take
$k_2=0.3~h{\rm Mpc}^{-1}$).  The idea is that to model the BAO wiggles
with a simple polynomial would require that the polynomial contains as
many maxima and minima as there are peaks and troughs in the wiggles.
Thus if we keep the fitting polynomial at a sufficiently low order the
wiggles will be automatically excluded from the fit.

A fourth order polynomial in principle has five free parameters, but
since we impose continuity in $k_1$ and $k_2$ together with a zero
derivative in $k_1$, we only need to adjust two free parameters for
each model.  This is done using a simple least-square fit.  We find in
several independent runs that this method gives the same parameter
estimates as the spectral method up to very high accuracy.

\subsection{Semi-analytic fitting formula}

The original fitting formula of Ref.~\cite{Eisenstein:1997ik} was
extended in Ref.~\cite{Eisenstein:1997jh} to allow for a non-zero
$f_\nu$, and optimised for the case of one massive and two massless
neutrinos, with $N_0+N_{\rm m}=N_{\rm eff}$ held fixed at three.  The
relative error is estimated at the $< 4$\% level.
Reference~\cite{Kiakotou:2007pz} further extended the work of
Ref.~\cite{Eisenstein:1997jh} by including $w$ as a free parameter.
This extension again assumed $N_{\rm eff}=3$, but was optimised
instead for $N_0=0$ and $N_{\rm m}=3$.  In the present work, we perform yet
another extension by relaxing the assumption of $N_{\rm eff}=3$, so
that $N_0$ and $N_{\rm m}$ are two completely independent free parameters.
Our code is built on that of Ref.~\cite{Kiakotou:2007pz}. 

In the following description of our modifications, however, we make
frequent reference to~\cite{Eisenstein:1997jh} since the equations are
better documented there.
\begin{enumerate}
\item{Equation~(1) of~\cite{Eisenstein:1997jh} gives the redshift of
    matter-radiation equality as
    \begin{equation}
      z_{\rm eq} = 2.50 \times 10^4 \  \Omega_{\rm m} h^2  \Theta_{2.7}^{-4}-1.
    \end{equation}
    We modify this expression to
    \begin{equation}
      z_{\rm eq} =  A(N_0,N_{\rm m}) \times 10^5 \  \Omega_{\rm m} h^2  \Theta_{2.7}^{-4} -1,
    \end{equation}
    where
    \begin{equation}
      A(N_0,N_{\rm m}) =  [2.32 + 0.56 (N_0+N_{\rm m})]^{-1}.
    \end{equation}
  \item Equation~(5)  of~\cite{Eisenstein:1997jh} introduces a normalised wavenumber $q$,
    \begin{equation}
      q \equiv  \frac{k}{{\rm Mpc}^{-1}} \Theta_{2.7}^2 (\Omega_{\rm m} h^2)^{-1} = 0.0746 \ k/k_{\rm eq},
    \end{equation}
    where
    \begin{equation}
      k_{\rm eq} = 0.0746 \ \Omega_{\rm m} h^2 \Theta_{2.7}^{-2}.
    \end{equation}
    Here, we replace the expression for $k_{\rm eq}$ with
    \begin{equation}
         k_{\rm eq} = 0.1492\,\sqrt{A \left(N_0 ,\,N_{\rm m}
         \right)}\,\Omega_{\rm m} h^2\; \Theta_{2.7}^{-2} \,.
    \end{equation}}
\item{The fitting coefficients in equation~(15)
    of~\cite{Eisenstein:1997jh}
    \begin{equation}
     \alpha_\nu (f_\nu,f_b,y_d) = \frac{f_c}{f_{cb}} \frac{5-2 \left( p_ c + p_{cb}
    \right) }{5 - 4 p_{cb}} \frac{(1-0.553f_{\nu b} + 0.126 f_{\nu
    b}^3 )}{1-0.193 \sqrt{ f_\nu N_{\rm m}} + 0.169 f_\nu N_{\rm m}^{0.2}} \times
    \ldots
    \end{equation}
    are optimised to
    \begin{equation}
      \alpha_\nu (f_\nu,f_b,y_d) = \frac{f_c}{f_{cb}} \frac{5-2 \left(
      p_c + p_{cb} \right) }{5 - 4 p_{cb}} \frac{(1-0.553f_{\nu b} +
      0.126 f_{\nu b}^3 )}{1-0.193 \sqrt{ f_\nu } N_{\rm m}^{0.2} + 0.169
      f_\nu} \times \ldots
    \end{equation}}
\item{Finally, we define
    \begin{equation}
      \Gamma_{\rm eff} = \sqrt{\alpha_\nu} + \frac{1-\sqrt{\alpha_\nu}}{1+ (0.43 ks)^4},
    \end{equation}
    so that
    \begin{equation}
      q_{\rm eff}=q/\Gamma_{\rm eff},
    \end{equation}
    cf.~equations~(16) and (17) of~\cite{Eisenstein:1997jh}.} 
\end{enumerate}
For models with $\{N_0=0, N_{\rm m}<6\}$, and $\{N_0<6,N_{\rm m}=3\}$, the fitting
formula is accurate to $<5$\% at $k < 0.2 \ h {\rm Mpc}^{-1}$

\section{BAO data in extended models}
\label{sec:appendix_bao}

Percival et al.~\cite{Percival:2009xn} give the BAO likelihood in the
form of a joint constraint on the parameters $d_{0.2}$ and $d_{0.35}$,
defined by
\begin{equation}
\label{eq:dz}
d_z = r_{\rm s} (\tilde{z}_{\rm d})/D_V(z),
\end{equation}
where
\begin{equation}
D_V(z) = \left[ (1+z)^2 D_A^2 c z/H(z) \right]^{1/3},
\end{equation}
with the angular diameter distance $D_A$, the Hubble parameter
$H(z)$, and
\begin{equation}
\label{eq:cosoho}
r_{\rm s} (z) = \int_0^{\eta(z)} {\rm d}\eta \; c_{\rm s} (1+z)
\end{equation}
is the comoving sound horizon.  Here, the sound speed is given by
\begin{equation}
c_{\rm s} = \frac{1}{\sqrt{3(1+R)}},
\end{equation}
where $R \equiv 3\rho_{\rm b}/4\rho_{\rm \gamma}$ is the ratio of
baryon to photon momentum density.

The sound horizon (Eq.(\ref{eq:cosoho})) should in principle be
evaluated at the baryon drag epoch $z_{\rm d}$, which is defined as
the redshift at which the drag optical depth $\tau_{\rm d}$ equals
one, i.e.,
\begin{eqnarray} \label{eq:zd}
  \tau_{\rm d}(\eta_{\rm d}) & \equiv & \int_{\eta}^{\eta_0} {\rm
    d}\eta' \; \dot{\tau}_{\rm d} \nonumber \\
  &=&  \int_{\eta}^{\eta_0} {\rm  d}\eta' \; \frac{\dot{\tau}}{R}
  \nonumber \\
  &=&  \int_{0}^{z_{\rm d}} {\rm  d}z \; \frac{{\rm d}\eta}{{\rm d}a} \,
  \frac{x_e(z) \, \sigma_{\rm T}}{R} \nonumber \\
  &=&  \frac{4}{3} \frac{\omega_\gamma}{\omega_{\rm b}} \int_{0}^{z_{\rm d}} {\rm  d}z \;
  \frac{{\rm d}\eta}{{\rm d}a} \, \frac{x_e(z) \, \sigma_{\rm T}}{1+z}
  = 1,
\end{eqnarray}
with today's baryon and photon densities $\omega_{\rm b}$,
$\omega_\gamma$, the Thomson cross-section $\sigma_{\rm T}$ and the
fraction of free electrons $x_e(z)$.  However, Percival et al.\ use
the approximation $\tilde{z}_{\rm d}$ obtained from a (somewhat
outdated) fitting formula (Eq.(4) of Eisenstein \&
Hu~\cite{Eisenstein:1997ik}) instead.  The fact that the results of
the fitting formula are inaccurate by several per cent is not a
problem {\it per se}, since $d_z$ is only used as a proxy in the data
processing, as discussed in Section~4 of Ref.~\cite{Percival:2009xn},
and as long as one considers simple cosmologies, the dependence of
$d_z$ on cosmological parameters is likely to be adequately
represented.

One should keep in mind though that the Eisenstein \& Hu fitting
formula was derived under certain assumptions (e.g., $N_{\rm eff}$ is
fixed to a value of 3). So for instance in models in which $N_{\rm
  eff}$ is a free parameter, the fitting formula will no longer
faithfully capture the proper parameter dependencies.  In these cases,
an unreflected use of the default BAO likelihood code as implemented
in the October 2009 version of \texttt{CosmoMC} will lead to biased
results.  However, this deficiency can be remedied by simple
rescaling.  Instead of using $d_z$ as defined in
Equation~(\ref{eq:dz}) as input for the BAO likelihood code, we
perform the substitution
\begin{equation}
  d_z \to  d_z \, \frac{\hat{r}_{\rm s}(\tilde{z}_{\rm d})}{\hat{r}_{\rm s}(z_{\rm
      d})} \,  r_{\rm s}(z_{\rm d}),
\end{equation}
where $\hat{r}_s$ is evaluated for the fiducial cosmology of
Ref.~\cite{Percival:2009xn}, and $r_{\rm s}(z_{\rm d})$ computed
numerically by solving Equation~(\ref{eq:zd}).  This quantity has the
correct dependence on all cosmological parameters and will yield
unbiased parameter and error estimates.


\end{document}